\documentclass[aps,prd,showpacs,nofootinbib,floats,floatfix,preprintnumbers,groupedaddress,twocolumn]{revtex4}

\usepackage{bm}
\usepackage{latexsym}
\usepackage{dcolumn}
\usepackage{amsmath,amsfonts,amssymb}
\usepackage{graphicx,epsfig}
\usepackage{amsmath}
\usepackage{fancyhdr}
\usepackage{hyperref}
\usepackage{graphicx,epstopdf}
\hypersetup{
	colorlinks   = true, 
	urlcolor     = blue, 
	linkcolor    = blue, 
	citecolor   = red 
}

{\scriptsize }

\def\p{\partial}

\begin{document}
\title{BMS Goldstone modes near the horizon of a Kerr black hole are thermal}
\author{Mousumi Maitra\footnote {\color{blue} maitra.91@iitg.ernet.in}}
\author{Debaprasad Maity\footnote {\color{blue} debu@iitg.ernet.in}}
\author{Bibhas Ranjan Majhi\footnote {\color{blue} bibhas.majhi@iitg.ernet.in}}

\affiliation{Department of Physics, Indian Institute of Technology Guwahati, Guwahati 781039, Assam, India
}

\date{\today}


\begin{abstract}
Near horizon Bondi-Metzner-Sachs (BMS) like symmetry is spontaneously broken by the black hole background itself and hence gives rise to Goldstone mode. The associated Goldstone mode for the near horizon BMS like symmetry of a Schwarzschild black hole was found to behave like inverted harmonic oscillators, which has been further shown to lead to thermodynamic temperature in the semi-classical regime. Here we investigate the generalization of this previous findings for the Kerr black hole. The analysis is being performed for two different situations. Firstly, we analyze Goldstone mode dynamics considering {\it slowly rotating} Kerr. In other case the problem is solved in the frame of {\it zero angular momentum observer} (ZAMO) with arbitrary value of rotation. In both analysis the effective semi-classical temperature of Goldstone modes turn out to be proportional to that of Hawking temperature. Due to such similarity and generality we feel that these Goldstone modes may play important role to understand the underlying microscopic description of horizon thermalization.     

\end{abstract}

\maketitle
\section{introduction}

 Thermodynamic properties of black holes is well studied but still an ill understood subject. It is by now a well established fact that symmetry could play crucial role in understanding the microscopic origin of such thermodynamic behavior of the black holes.  
In this regard asymptotic symmetries near the horizon attracted widespread interest in the recent time. Such symmetries are well known as Bondi-Metzner-Sachs (BMS) symmetry \cite{Bondi:1962px}-\cite{Akhmedov:2017ftb}. Those are the symmetry transformation which acts non-trivially on the states near the null boundary of an asymptotically flat black holes. The original study \cite{Sachs:1962zza} was done on the asymptotic null boundary of an asymptotically flat  spacetime. Subsequently the analysis has been generalized for arbitrary null boundary embedded in the black hole spacetime \cite{Cai:2016idg}-\cite{Akhmedov:2017ftb}. Symmetries near the horizon which is a special null boundary is believed to play crucial role in shedding light on the microscopic understanding of the black hole thermodynamics. 
It has already been established that conserved quantities of a special class of such symmetry can be associated with the horizon entropy of a black hole \cite{Iyer:1994ys}-\cite{Majhi:2015tpa}. Then after volume of works have been performed to inculcate the precise role of these symmetries in revealing the microscopic origin of the horizon thermodynamics \cite{Strominger:1997eq}-\cite{Setare:2016qob}.

It has been argued recently \cite{Dvali:2011aa}-\cite{Eling:2016qvx} that such assymptotic symmetries are spontaneously broken by the background spacetime. Therefore, there exists associated Goldston modes which are conjectured to be connected with the soft hair of the black hole spacetime.  Furthermore, interesting conjecture between the black hole microstates and the aforementioned soft hairs have been explored in  \cite{Dvali:2011aa}-\cite{Hawking:2016msc}.
Following these line of arguments, in our previous work \cite{Maitra:2019eix}, we proposed a Lagrangian dynamics of those Goldston modes emerging due to spontaneous breaking of the aforementioned near horizon BMS symmetries by the vacuum Schwarschild black hole. Dynamics of the Goldstone modes have been shown to behave like inverse harmonic oscillators. Quantization of those inverse harmonic oscillators were found to have interesting connection with the thermal properties of the black hole. Following conjecture proposed in \cite{Morita:2018sen}\cite{Morita:2019bfr} and also  \cite{Maldacena:2015waa}-\cite{Bombelli:1991eg}, we were able to define a thermodynamic temperature which turned out to be proportional to the Hawking temperature of the Schwarzschild black hole (see also \cite{Hegde:2018xub,Dalui:2019esx,Dalui:2020qpt,Subramanyan:2020fmx,Majhi:2021bwo,Dalui:2021tvy} for connecting inverse harmonic oscillator with thermalization). The whole idea  has been successfully extended to black hole solution in three dimensional massive gravity \cite{Setare:2021gzm}. 

In the present paper, we generalize our previous construction for the Kerr black hole. Our current study not only shed light on the effect of the black hole spin, but also can be thought of as a first non-trivial check of our earlier proposal \cite{Maitra:2019eix} of physics of Goldstone modes in the gravitational sector. 
In this case also we observe that the dynamics of Goldstone modes associated with the super-translation parameter satisfy harmonic oscillator equation with spin dependent inverted potential. This inverted harmonic oscillator dynamics has been further conjectured to lead to the quantum mechanical temperature which turned out to be proportional to Hawking temperature. We have performed our present analysis for two different cases. One is for a slowly rotating black hole with arbitrary observer. For second case we compute the temperature for a special class of Zero angular momentum observers (ZAMO) for arbitrary spin parameter. Our present analysis reconfirms the fact that the BMS like symmetries may play important roll to understand the microscopic origin of black hole thermodynamics.

%
\section{Near horizon symmetries and modified metric}
In order to investigate the BMS like symmetry near the horizon, the suitable form of the metric is usually converted in Gaussian null coordinates (GNC) near the horizon. However the derivation of the Kerr metric in GNC  has been explicitly presented in \cite{Booth:2012xm} and also near horizon symmetry vector has been derived in details in the literatures \cite{Donnay:2015abr, Akhmedov:2017ftb}. Nevertheless below we present a brief description of the derivations as those results are important and also useful for the subsequent part of the present analysis.  
In Eddington-Finkelstein coordinate Kerr metric is written as \cite{Booth:2012xm},
\begin{eqnarray}
&&ds^2 = -(1-\frac{\Delta -\chi}{\Sigma}) dv^2 +2 dv dr \nonumber
\\
&& - \frac{2 a (\chi-\Delta) \sin^2\theta}{\Sigma} dv d\phi- 2 a \sin^2\theta dr d\phi
\nonumber
\\
&&+ \Sigma d\theta^2+ \frac{\sin^2\theta [\chi^2 -a^2 \sin^2\theta \Delta]}{\Sigma}d\phi^2~, 
\label{kmetric}
\end{eqnarray}
where $\Delta= r^2+a^2 -2 M r;~  \chi= r^2+a^2;~ \Sigma= r^2+a^2 \cos^2\theta$.
It is well known that within ergosphere of the Kerr black hole observers cannot be static, rather  they must feel frame-dragging effect and corotate with the black hole because of the presence of $g_{v\phi}$ term in the metric. So in order to have reasonable physical picture of the near horizon geometry in stationary rotating background, one has to introduce a coordinate transformation to the dragging frame as given by,
\begin{eqnarray}
v \rightarrow v,~~ r \rightarrow r,~~~\theta \rightarrow \theta,~~~\phi \rightarrow \phi + (a/ \chi_H) v.
\end{eqnarray}
By this transformation $g_{v\phi}$ vanishes on the horizon of the Kerr metric. In this coordinate system, null vector $\frac{\partial}{\partial v}$ will be orthogonal to the surfaces of $v=$constant. The transformed metric becomes,
\begin{eqnarray}
&&ds^2 = -(\frac{\Sigma^2_H \Delta -a ^2 \sin^2\theta (r-r_H)^2}{\Sigma \chi^2_H}) dv^2 +2 \frac{\Sigma_H}{\chi_H} dv dr\nonumber\\
&&+ 2 a \sin^2\theta \frac{\Delta \Sigma_H + \chi (r^2-r^2_H)}{\Sigma (r^2_H +a^2)} dv d\phi- 2 a \sin^2\theta dr d\phi \nonumber
\\
&&+ \Sigma d\theta^2 + \frac{\sin^2\theta [\chi^2 -a^2 \sin^2\theta \Delta]}{\Sigma}d\phi^2~. 
\label{kmetric1}
\end{eqnarray}

Now we will follow standard procedure (presented in \cite{Booth:2012xm}) to construct near horizon metric in Gaussian null coordinate system. For this reason we have to define the suitable pair of null normals on the horizon as,
\begin{eqnarray}
l^a=(1,0,0,0); \,\,\,\ n^a=(\frac{a^2 \sin^2\theta}{2 \Sigma_H}, \frac{\chi_H}{\Sigma_H},0, \frac{a}{\chi_H}).
\end{eqnarray}
These normals are defined such that at the horizon $ n^a n_a=0; l^a l_a=0; $ and $l^a n_a=1$.

Next we consider a set of incoming null geodesics which crosses the horizon having tangent vector $n^a$. So the geodesics can be parametrized with affine parameter $\rho$ ($\rho\approx r-r_H$) such that  the horizon surface is identified as $\rho=0$, but increases towards the inside. Therefore the  null geodesics curve $X^a (\rho)$ with $X^a = (v,r,\theta,\phi)$, can be expanded upto second order of $\rho$ as follows:
\begin{eqnarray}
X^a (\rho) \approx   X^a \vline_{\rho= 0} +\rho \frac{dX^a}{d\rho} \vline_{\rho=0} + \frac{{\rho}^2}{2} \frac{d^2 X^a}{d\rho^2} \vline_{\rho= 0} +\dots~.
\label{geodesic}
\end{eqnarray}
In the right hand expression the components of first term  is identified as $X^a \vline_{\rho=0} =( v,r_H,\theta, \phi)$. The second one can be expressed as the tangent vector to the curve, $\frac{dX^a}{d\rho} \vline_{\rho=0} = n^a\vline_{\rho=0}$. 
The third term can be written with the help of null geodesics equation at the horizon as,
\begin{eqnarray}
\frac{d^2 X^a}{d \rho^2} \vline_{\rho=0} = -\Gamma^a_{b c} n^b n^c\vline_{\rho=0}.
\end{eqnarray} 
Using all these results in Eq. (\ref{geodesic}) and keeping upto first order in $\rho$, we can define a  transformation of coordinates from $(v',r,\theta',\phi')$ to $(v,\rho, \theta,\phi)$ in the following way,
\begin{eqnarray}
&& v = v' + \rho (\frac{a^2 \sin^2\theta}{2 \Sigma_H});\,\,\,\ r= r_H + \rho  (\frac{\chi_H}{\Sigma_H})~;
\\
\nonumber
&& \theta=  \theta';\,\,\,\  \phi = \phi' +\rho (\frac{a}{\chi_H})~.
\label{tranf}
\end{eqnarray}
So by having tensor transformation rule of the components, we can obtain the metric in Gaussian null coordinate system $(v, \rho, \theta,\phi)$ as follows,
\begin{eqnarray}
&& ds^2 =-2 \rho \kappa dv^2+2 dv d\rho +2 \rho \beta_A dv dx^A +(\mu_{AB}\nonumber
\\
&& + \rho \lambda_{AB}) dx^A dx^B~.
\label{kerr} 
\end{eqnarray}
where the corresponding quantities in the metric are defined by;
\begin{eqnarray}
&&\kappa= \frac{\Delta'(r_H)}{2 \chi_H};~~~\beta_{\theta}=\frac{2 a^2 \sin\theta \cos\theta}{\Sigma_H};\nonumber
\\
&& \beta_{\phi}=\frac{a \sin^2\theta}{\Sigma_H}\Delta'(r_H)+\frac{2 a r_H \chi_H \sin^2\theta}{\Sigma^2_H}\nonumber
\\
&&\mu_{\theta\theta}= \Sigma_H;~~~ \mu_{\phi\phi}= \frac{\chi^2_H \sin^2\theta}{\Sigma_H}~; ~~~~\lambda_{\theta\theta}=\frac{2 r_H \chi_H}{\Sigma_H}~;\nonumber
\\
&& \lambda_{\theta\phi}= \frac{2 a^3 \chi_H \sin^3\theta \cos\theta}{\Sigma^2_H};
\nonumber
\\
&& \lambda_{\phi\phi}= \Big(-\frac{a^2 \chi_H \sin^4\theta}{\Sigma^2_H} \Delta'(r_H) \nonumber
\\
&& + \frac{2 r_H \chi^2_H \sin^2\theta (\Sigma_H -a^2 \sin^2\theta)}{\Sigma^3_H}\Big).  
 \end{eqnarray}
Here the horizon is located at $\rho=0$. 
To obtain the asymptotic symmetries near the horizon, the appropriate fall-off conditions for the metric coefficients are assumed to be 
\begin{eqnarray}
\pounds_\zeta  g_{\rho \rho}= 0, \ \ \ \pounds_\zeta  g_{v \rho}=0, \ \ \ \pounds_\zeta  g_{A \rho}=0~;\label{con}\\
\pounds_\zeta  g_{vv} \approx \mathcal{O}(\rho); \ \ \ \pounds_\zeta  g_{vA} \approx \mathcal{O}(\rho); \ \ \  \pounds_\zeta  g_{AB} \approx \mathcal{O}(\rho)~,
\label{con1}
\end{eqnarray}
such that the transformed metric assume the same form near the horizon $\rho = 0$.
Here, $\pounds_\zeta$ corresponds to the Lie variation associated with the diffeomorphism $x^a\rightarrow x^a+\zeta^a$ and is given by 
\begin{eqnarray}
\pounds_{\zeta} g_{ab} = \nabla_a\zeta_b+\nabla_b\zeta_a\label{LieK}
\end{eqnarray}

Following the gauge choices given in Eq.(\ref{con}), the components of the diffeomorphism vector can be expressed as,
\begin{eqnarray}
\zeta^v &=& F(v,x^A)~;\nonumber\\
\zeta^{\rho} &=& T(v,x^A)- \rho \partial_v F- \partial_B F \int \rho \beta^{B} d\rho~;\nonumber\\ 
\zeta^A &=& - \partial_B F \int \mu^{AB} d\rho + R^A(v,x^A)~.
\label{diff}
\end{eqnarray}
Now we impose the weak fall off conditions given in (\ref{con1}), on the aforementioned solutions (\ref{diff}). From there  the following constraints relations have appeared among the diffeomorphism parameters $T$ and $R^A$,
\begin{eqnarray}
&&\partial_v T + \kappa T= 0~;\label{con3}
\\
&&\partial_A T - T(v,x^A) \beta_A 
+ \mu_{AB} \partial_v  R^B  =0~; \label{con4}
\\
&& R^D \partial_D \mu_{AB} + \mu_{AD} \partial_B R^D + \mu_{BD} \partial_A R^D = 0~.\label{con5}
\end{eqnarray}
The equation (\ref{con5}) boils down to the three component equations as,
\begin{eqnarray}
&& R^{\theta} \partial_{\theta} \mu_{\theta\theta} + 2 \mu_{\theta \theta} \partial_{\theta} R^{\theta}=0;~~~\nonumber
\\
&& \mu_{\theta \theta} \partial_{\phi} R^{\theta} + \mu_{\phi \phi} \partial_{\theta} R^{\phi}=0;\label{con5_1}
\nonumber
\\
&& R^{\theta} \partial_{\theta} \mu_{\phi \phi} +2 \mu_{\phi \phi} \partial_{\phi} R^{\phi}=0~.
\end{eqnarray}
Now, we solve the above equations with the condition that the position of the null surface at $\rho=0$ remain unaltered which leads to $T(v,x^A)=0$ from Eq.(\ref{diff}). Then (\ref{con4}) yields that $R^A$ is independent of $v$. Furthermore, the leading order term in $\beta^A$ is independent of $\rho$, and the last term in $\zeta^\rho$ yields sub-leading $\mathcal{O}(\rho^2)$ contribution.
With this, the diffeomorphism vector $\zeta^a$ corresponding to above conditions assumes the following form \cite{Donnay:2015abr, Akhmedov:2017ftb}
 \begin{eqnarray}
\zeta^a \partial_a=&& F(v,x^A) \partial_v - \rho  \partial_v F(v,x^A) \partial_{\rho}
\nonumber
\\ 
&+& \big(-\rho  \partial^{A} F(v,x^A) + R^A(x^A)\Big) \partial_A~.
\label{zeta}
\end{eqnarray}

Near the horizon $\rho=0$ surface we have two diffeomorphism parameters $F$ and $R^A$ associated with \emph{ supertranslation} and \emph{ superrotation} symmetry parameter respectively \cite{Donnay:2015abr, Akhmedov:2017ftb, Maitra:2018saa}. 
Under this asymptotic symmetry transformation, the corrected metric is expressed as $g_{ab}= g^{(0)}_{ab} +h_{ab}$, where $g^{(0)}_{ab}$ is given by Eq.(\ref{kerr}) and the components of $h_{ab}$ are derived by the Lie derivative of the metric as given by (\ref{LieK}); i.e. $h_{ab}= \pounds_{\zeta} g_{ab} = \nabla_a\zeta_b+\nabla_b\zeta_a$. So these corrected components are expressed up to linear order by the symmetry transformations as follows,
\begin{eqnarray}
&&h_{ab}dx^a dx^b=
- [2 \rho (\kappa \partial_v F +\partial^2_v F)] dv^2 \nonumber
\\
&+&2 \rho [R^B \partial_B \beta_A -2 \kappa \partial_A F -\partial_A \partial_v F \nonumber
\\
&+& \beta_B \partial_A R^B -\mu_{AB} \partial_v \partial_D F \mu^{ BD}]dv dx^A\nonumber
\\
& +& \rho  \Big[-\lambda_{AB} \partial_v F -\partial_E F \mu^{ DE} \partial_D \mu_{AB} \nonumber
\\
&+& \beta_A \partial_B F + \beta_B \partial_A F - \mu_{AD} \partial_B(\partial_E F  \mu^{ DE})\nonumber
\\
&-&\mu_{BD} \partial_A(\partial_E F  \mu^{ DE}) + R^D \partial_D \lambda_{AB}\nonumber
\\
& + & \lambda_{AD} \partial_B R^D +\lambda_{BD} \partial_A R^D  \
\Big] dx^A dx^B~.
\label{corrected}
\end{eqnarray}
The corrected metric can be understood as the change of physical parameters of the background black hole with the following transformation,
\begin{eqnarray}
&&\kappa \rightarrow \kappa + \kappa \partial_v F +\partial^2_v F; \nonumber
\\
&&\beta_A \rightarrow \beta_A + R^B \partial_B \beta_A -2 \kappa \partial_A F -\partial_A \partial_v F 
\nonumber
\\
&&+ \beta_B \partial_A R^B -\mu_{AB} \partial_v \partial_D F \mu^{ BD};\nonumber
\\
&& \lambda_{AB} \rightarrow  \lambda_{AB} + \Big(-\lambda_{AB} \partial_v F -\partial_E F \mu^{ DE} \partial_D \mu_{AB}\nonumber
\\
&& + \beta_A \partial_B F + \beta_B \partial_A F - \mu_{AD} \partial_B(\partial_E F  \mu^{ DE})\nonumber
\\
&& -\mu_{BD} \partial_A(\partial_E F  \mu^{ DE}) + R^D \partial_D \lambda_{AB}\nonumber
\\
&&+ \lambda_{AD} \partial_B R^D +\lambda_{BD} \partial_A R^D \Big).\label{corr}
\end{eqnarray}
These corrections of the macroscopic parameters of the background geometry by the asymptotic symmetry transformation can be thought of as the symmetry breaking phenomena near the horizon \cite{Eling:2016xlx}.
It is well known  that the spontaneous breaking of the global continuous symmetry can lead to massless Goldstone bosons. However the breaking of gauge symmetry does not give Goldstone particles. The underline symmetry in gravity theory is the diffeomorphism which is a set of infinitesimal local coordinate transformation of spacetime. Therefore diffeomorphism can be interpreted to be gauge symmetry in gravity. However bulk diffeomorphism can act as the global symmetry on the boundary surface. Here we discuss about the asymptotic symmetries which is actually the global symmetry arising at the horizon boundary. Hence we can say that the Goldstone Boson should appear due to the spontaneous breaking of this boundary global symmetry. In the present analysis at first we have identified those specific set of diffeomorphism under which the asymptotic structure of the metric remains intact. Nevertheless in this process the black hole solution is transformed and physical parameters of the black hole are modified. For instance the parameter  $\kappa$  get shifted to $\kappa'$ keeping intact the near horizon boundary conditions. Strictly speaking the above phenomena does not show the breaking of the near horizon symmetries. However this shows that among all possible solutions of Einstein equations, characterised by arbitraryness of $F$ and $R^A$, Kerr is only one choice.  Hence the background Kerr solution breaks the symmetry of considering all possible values of parameters $F$ and $R^A$. This nature is identical to spontaneous symmtery breaking by the vacuum expectation value (VEV) of fields in QFT, if we think the choice as Kerr is similar to VEV in QFT.  Therefore the parameter $F$ which characterize this symmetry breaking, is promoted as the Goldstone boson modes.


Furthermore, in the Hamiltonian formulation we know $\kappa$ is associated with the Lagrangian multipliers of the theory, which are usually kept fixed everywhere in the spacetime \cite{Eling:2016xlx}-\cite{Eling:2016qvx}. Then appropriate boundary terms has to be added with action such that on-shell variation vanishes. With this condition on $\kappa$, the Eq.(\ref{corr}) is solved for $F$, which becomes non-dynamical. However, if we impose more general condition on $\delta \kappa$ being non-vanishing everywhere in the bulk except at the horizon boundary (denoted by $r=r_H$ or $v\to -\infty$), satisfying
 \begin{equation}
 \lim_{v\rightarrow-\infty} \Big(\alpha \partial_{v} F + \partial^2_v F\Big) =0, \label{boundaryke}, 
 \end{equation}	
 the super-translation parameter $F$ becomes dynamical in nature.
Therefore keeping this is mind, we first construct an appropriate Lagrangian of $F$, and pick those solutions of $F$ which will naturally satisfy Eq. (\ref{boundaryke}). To this end let us point out that  in some literatures \cite{Bunster:2014mua} \cite{Perez:2016vqo} the fluctuation of the Lagrangian multiplier has been considered even on the boundary but then also correct boundary terms has to be added with the action. We have not discussed this part in the present manuscript.

 It may be pointed out that $R^A$ does not depend upon timelike coordinate $v$, and $F$ depends on $v$. Consequently $F$ will be the physical dynamical Goldstone mode for Kerr background. The rotational parameter will be non-dynamical in nature and that must be consistent with the three constraint conditions expressed in  Eq.(\ref{con5_1}).
 

Therefore, consistent with the constraints as discussed above, simplest possible choice for $R^A$ is 
\begin{eqnarray}
	R^{\theta}=0;~~~~~~~R^{\phi}=C = \textrm{constant}~. \label{R}
\end{eqnarray}
  We will later see that such choice makes $\delta \beta_A$ to vanish automatically at the horizon. This is reminiscent to the condition $\delta\kappa = 0$ at the horizon boundary. 
The constraint relations in (\ref{con5_1}) impose over restriction on parameters $R^{\theta}$ and $R^{\phi}$. The above solution Eq.(\ref{R}) the simplest one which will satisfy all the conditions. There may exist other solution, however, for our present purpose we restrict to this specific choice.

\section{\label{lag} Lagrangian and equation of motion}
In this section our aim is to construct dynamical equations of the two Goldstone modes with the aforementioned constraint conditions on the black hole parameters. Following the methodology formulated in \cite{Maitra:2019eix}, appropriate dynamical Lagrangian is conjectured to have two terms. Important term is the Einstein-Hilbert Lagrangian corresponding to the corrected metric $g_{ab}$ near the black hole horizon $\rho=0$,
\begin{eqnarray}
\mathcal{L}_{\mathcal{F,R^A}} =  \sqrt{-g'} \mathcal{R'}~. 
\label{L}
\end{eqnarray}
$g'$ and $\mathcal{R}'$ are the determinant and Ricci scalar of the corrected metric $g_{ab}$.
Here  the modified metric $g_{ab}$ has been constructed by taking into account a subset of diffeomorphism which preserve the near horizon boundary conditions. It is well known that due to these diffeomorphism, the Einstein-Hilbert Lagrangian must be invariant upto a total derivative terms which also become boundary term  on the closed boundary enclosing a bulk region of the manifold. However in this analysis we have analyzed near the horizon $(r=0)$ which is a part of this closed boundary. Therefore this total derivative term will have finite contribution on this part of the boundary. The form of the Lagrangian presented in the main text comes from this non-zero contribution (more detail can be found in the reference \cite{Maitra:2019eix}, Section IIA after equation (16)). Now  the solution of the Einstein equations is modified having a transformed form of the black hole parameters where the transformation is given by first order derivative of $F$. The Einstein's equation of motion are derived from the variation of $g_{ab}$. It may be noted that in the case of modified metric, $g_{ab}$ is now not simply $F$, rather it is a non-trivial function of $F$. Therefore the dynamical equation of motion for $F$ cannot be derived from  the Einstein equations as it is not picking the exact flavor of variation of $F$. Hence in order to find a equation of $F$, it is always necessary to find an action for this parameter, which we have done in the subsequent analysis.

 Now to calculate above Lagrangian we follow the procedure usually adopted in the context of stretched horizon fluid dynamics (detail discussion can be found in section 4 of \cite{Carlip:1999cy}). We first calculate ({\ref{L}}) on the radial coordinate $\rho=\epsilon$ very near the horizon, then $\epsilon \rightarrow 0$ limit will be taken to get the required expression of the Goldstone boson Lagrangian.

Now  we can expand the above Lagrangian in Taylor series with respect to the background metric $g^0_{ab}$ (\ref{kerr}), assuming $h_{ab}$ as the small fluctuation. The expansion of the Lagrangian is given by,
\begin{eqnarray}
&&\mathcal{L_{F,R^A}} = \mathcal{L_{F,R^A}}(g_{ab}^{(0)}) + h_{ab}\Big(\frac{\delta\mathcal{L_{F,R^A}}}{\delta g_{ab}}\Big)^{(0)} \nonumber 
\\
& &+  h_{ab}h_{cd}\Big(\frac{\delta^2\mathcal{L_{F,R^A}}}{\delta g_{ab} \delta g_{cd}}\Big)^{(0)}+\dots~.
\label{L1}
\end{eqnarray}
As has been explained in our previous paper \cite{Maitra:2019eix}, the free Lagrangian for the Goldstone mode will be the third term which is quadratic in the parameters. For the present purpose we will restrict ourself only up to second order in Goldstone modes. Using the constraints on the super-rotation parameter in (\ref{R}),
 %
the Lagrangian for the super-translation modes becomes,   
\begin{eqnarray}
&&\mathcal{L}_{(F,R^{\theta}=0,,R^{\phi}=C)}= c_1(\theta) (\partial_v F)^2 +c_2(\theta) (\partial_{\theta} F)^2 \nonumber
\\
&+& c_3(\theta) (\partial_{\phi} F)^2 + c_4(\theta) (\partial_v \partial_{\theta} F)^2 +c_5(\theta) (\partial_v \partial_{\phi} F)^2 \nonumber
\\
&+& c_6(\theta) \partial_{\theta}F \partial^2_v F + c_7(\theta) \partial_{\phi} F\partial^2_v F +  c_8(\theta) \partial^2_{\theta}F \partial^2_v F \nonumber \\
&+& c_9(\theta) \partial^2_{\phi} F\partial^2_v F .\label{LF1}
\end{eqnarray} 
See Appendix A for a detail calculation explicit forms of $c_i$s. 
As has already been described in detail in our previous work \cite{Maitra:2019eix}, the Lagrangian is so constructed that the action from the modified metric $g_{ab}$ will describe the Goldstone mode dynamics near the the rotating black hole horizon.
Another part of the proposed Lagrangian is the Gibbons-Hawking-York (GHY) boundary term which is required to have consistent variational principle and is given by,
\begin{equation}
\mathcal{S}_2 =- \frac{1}{8\pi G}\int d^3x \sqrt{h}K~.
\label{GHY}
\end{equation}
 In the above expression the quantity $K$ is given by $K = -\nabla_a M^a$, which is the trace of the extrinsic curvature of the boundary surface (at $\rho \rightarrow 0$). Here $M^a$ is considered as the unit normal to the $\rho=constant$ hyper-surface. In Appendix B we showed that it gives a total derivative term and therefore does not contribute to the dynamics.  
  
In this scenario one important point remains left to be discussed here. As we discussed before that the Lagrangian is calculated using stretched horizon approach by cosidering a timelike surface very near from the horizon. So one may consider GHY term as the correct boundary term to have a proper variational priciple of the action. However it has been shown in \cite{Parattu:2015gga}\cite{Lehner:2016vdi} that if one want to perform any physical analysis exactly on the null surface, then from the first principle one can prescribe a well defined counter term on this surface. Therefore it will be interesting to study the present analysis considering the null boundary counter term.


Once we have the appropriate Lagrangian expressed in Eq.(\ref{LF1}), the equation of motion for $F$ yields as follows,
 \begin{widetext}
 \begin{eqnarray}
 && 12 \alpha_2^2 \Sigma_H  \Big( \Sigma_H \chi_H^2 \partial^2_{\theta} F +(r^2_H+2 a^2-a^2\cos^2\theta) \chi_H^2 (\cot\theta) \partial_{\theta} F+ \frac{\Sigma^3_H}{\sin^2\theta} \partial^2_{\phi} F \Big) -16 \chi^2_H \Sigma_H \Big( \Sigma_H \chi^2_H \partial^2_v \partial^2_{\theta} F \nonumber
 \\
 && + (r^2_H+2 a^2-a^2\cos^2\theta) \chi^2_H  (\cot\theta) \partial^2_v \partial_{\theta} F + \frac{\Sigma^3_H}{\sin^2\theta}\partial^2_v \partial^2_{\phi} F- \alpha \partial^2_v F \Big) =0~. 
 \label{kerrF}
 \end{eqnarray}
 Here $\alpha$ is given by
 \begin{eqnarray}
 &&\alpha= -8 r^3_H \Big(a^2 \alpha_1 \alpha_3  -4 M r_H^3 \alpha_2 \Big) + 8 a^2\Big( 8 a^2 \chi_H (7 r^2_H +3 a^2) + r^3_H \alpha_1 \alpha_3 + 2 a^2 r^2_H \alpha_2^2 \Big) \cos^2\theta\nonumber
 \\
 &&+ 8 a^4 \Big(-3  \chi^2_H +4 M r_H  \alpha_3 + \alpha_2^2  (a^2-2 r_H^2) \Big)\cos^4\theta -8 a^6 \alpha_2^2  \cos^6\theta.
 \end{eqnarray}
 The forms of the $\alpha_i$s are given in Appendix A.
 \end{widetext}
 
However for general value of background angular momentum parameter $a$, it turned to be very difficult to solve for the mode function $f_{lm}$. Hence, in the next section we will solve the problem for slowly rotating black holes. Later we will discuss the same for Zero angular momentum observers (ZAMO).


\subsection{Slowly rotating Kerr spacetime}
We consider slowly rotating background  where rotation parameter $a$ is such that $a \ll M$.  Considering $a/M =x$, we can expand (\ref{kerrF}) upto $\mathcal{O} (x^2)$ as follows (detail can be found in Appendix D and also in  Appendix E1) :
\begin{eqnarray}
&&    \Big[64 x^2 \sin^2\theta \p^2_{\theta} \p^2_v F + \Big( 12  \cos^2\theta  +84   \Big) x^2 \p^2_v F \nonumber
\\
&& + 128 x^2 \p^2_v \p^2_{\phi} F - 64 (2-3 x^2) \Big( \p^2_{\theta}\p^2_v F +\cot\theta \p_{\theta} \p^2_v F \nonumber
\\
&& +\frac{1}{\sin^2\theta} \p^2_{\phi} \p^2_v F \Big)\Big]\nonumber
 \\
 && - \frac{1}{M^2} \Big[ 3 x^2 \sin^2\theta \p^2_{\theta} F + 6  x^2  \p^2_{\phi} F- 6 (1-2 x^2) \Big( \p^2_{\theta} F \nonumber
 \\
 && +\cot\theta \p_{\theta} F +\frac{1}{\sin^2\theta} \p^2_{\phi} F \Big) \Big]=0\nonumber\\ \label{slow2}
\end{eqnarray}

In the original stationary axially symmetric background, the angular part of field  must satisfy the equation for spheroidal harmonics \cite{Brill:1972xj}. However in the present analysis because of the fluctuation around the background, the equation of $F$ does not satisfy the spheroidal harmonics equation. But still we have considered the angular part of the solution ansatz in the form of spherical harmonics. The validity of this choice is clear in this section as we have focussed in slowly rotating spacetime. Here we will try to find the dynamics of $F$  by the perturbative analysis where the zeroth order solution is the result found on spherically symmetric Schwarzschild background. Therefore in slowly rotating background the final solution of $F$ has been found out as the first order correction to that zeroth order solution. So we assume that the separability assumption in terms of spherical harmonics should work in the present analysis.
 We, therefore, expand field $F$ in terms of the generic spherical harmonic basic,
 \begin{eqnarray}
 F(v,\theta,\phi) = \sum_{l,m}  f_{lm}(v) Y_{lm}(\theta,\phi).\label{solF}
 \end{eqnarray}
  Putting solution ansatz (\ref{solF}) in the equation (\ref{slow2}), we get,
 \begin{eqnarray}
&& 4 (\p^2_v f_{lm}) \Big[ 16 x^2 \sin^2\theta (\p^2_{\theta} Y_{lm}) + \Big( 32 l(l+1) -16
\nonumber
\\
&& - x^2(48 l(l+1) -3 \cos^2\theta +32 m^2 ) \Big) Y_{lm} \Big] \nonumber 
\\
&&- \frac{3}{M^2} f_{lm}(v) \Big[ x^2 \sin^2\theta (\p^2_{\theta} Y_{lm}) 
\nonumber
\\
&&+ 2 \Big( l(l+1)  - x^2(2 l(l+1)+ m^2)\Big)  Y_{lm}  \Big]=0. \label{slowly}
\end{eqnarray}
The mode $l=0$ and $m=0$, satisfies,
\begin{eqnarray}
\p^2_v f_{00}=0 ,\label{l0}
\end{eqnarray}
whose solution is $f_{00}= b_1(x^A) + b_2(x^A) v$. The boundary condition $\delta\kappa=0$ at $v\to -\infty$ implies $b_2 = 0$. Therefore, the final solution will be $ f_{00}(v)=b_1$.
For generic mode $f_{lm}$ the equation (\ref{slowly}) can be expressed as,
\begin{eqnarray}
 &&\Big((2l(l+1)-1) \p^2_v f_{lm}-\frac{3 l(l+1)}{32M^2} f_{lm} \nonumber
 \\
 &&+ x^2  \Big[ (\beta_1  \p^2_v f_{lm} + \beta_2  f_{lm}) 
 \nonumber
 \\
 &&+ \sum_{m'=-l'}^{l'} \sum^{\mid 2+l \mid}_{l'= \mid 2-l \mid} (\beta_{l'm'lm} \p^2_v f_{l'm'}  
 \nonumber
 \\
 &&+  \bar{\beta}_{l'm'lm} f_{l'm'}) \Big] \Big) Y_{lm} =0; 
 \label{Y0}
 \end{eqnarray}
 See Appendix E2 for details.
Hence, we now have generic coupled mode equation,
 \begin{eqnarray}
 &&\Big((2l(l+1)-1) \p^2_v f_{lm}-\frac{3 l(l+1)}{32M^2} f_{lm} \nonumber
 \\
 &&+ x^2  \Big[ (\beta_1  \p^2_v f_{lm} + \beta_2  f_{lm}) 
 \nonumber
 \\
 &&+ \sum_{m'=-l'}^{l'} \sum^{\mid 2+l \mid}_{l'= \mid 2-l \mid} (\beta_{l'm'lm} \p^2_v f_{l'm'}  +  \bar{\beta}_{l'm'lm} f_{l'm'}) \Big] \Big) =0; \nonumber\\
 \label{Y1}
 \end{eqnarray}
We solve the aforesaid equation by using perturbative method in terms of rotation parameter $x$. The zeroth order part of the equation will correspond to that of the Schwarzschild background \cite{Maitra:2019eix}. Therefore, the solution is taken as, (for $ l \geq 1 $)
 \begin{eqnarray}
  f_{lm}=f^{sc}_{lm} +x^2 f^{(1)}_{lm} + \dots,\label{perturbsol}
 \end{eqnarray}
where,  $f^{sc}_{lm}$ is the form of the solution of $f(v)$ in Schwarzschild background given by \cite{Maitra:2019eix},
\begin{eqnarray}
f^{sc}_{lm}= \frac{c^{(0)}_{sc}}{k_{sc}} \exp[\Omega(l) \kappa_{sc} v],~~ \Omega(l)= \sqrt{\frac{3l(l+1)}{2 (2l(l+1)-1)}}. \nonumber\\
\label{Ksc}
\end{eqnarray}
Here $\kappa_{sc}$ is the surface gravity of Schwarzschild black hole and  $c^{(0)}_{sc}$ is a dimensionless constant. Hence $f^{sc}_{lm}$ have dimension of length. $f^{(1)}_{lm}$ is the leading order contribution due to slowly rotating Kerr black hole. The form of the equation $f^{(1)}_{lm}$ is obtained in Appendix F and according to the discussion presented there the equation of motion of each mode $f_{lm}(v)$ can be effectively expressed up to order $\mathcal{O}(x^2)$ as( for $l \geq 1$),
\begin{eqnarray}
\p^2_{v} f_{lm} - \kappa^2_{sc} (\Omega^2 + b x^2)  f_{lm} = 0. \label{f}
\end{eqnarray}
This solution contains already the zeroth order solution corresponding to  Schwarzschild background Eq.(\ref{Ksc}).
Now the solution of $f_{lm}$ will be,
\begin{eqnarray} 
f_{lm} &=& c^{(1)}_{kerr} \exp{[ (\sqrt{\Omega^2 +b x^2})\kappa_{sc} v]} \nonumber
\\
&+& c^{(2)}_{kerr} \exp{[- (\sqrt{\Omega^2 +b x^2})\kappa_{sc}v]}.\label{Fsol}
\end{eqnarray}
Here  $c^{(1)}_{kerr}$ and $c^{(2)}_{kerr} $ are the two undetermined constants of integration having dimension of length. In the above solution second term is diverging near the horizon boundary at $v\rightarrow -\infty$, and hence we set $c^{(2)}_{kerr}=0$.
Therefore the required solution turns into,
\begin{eqnarray}
f_{lm}= c^{(1)}_{kerr}  \exp{[\Omega(1+ \frac{b(l,m) x^2}{2 \Omega^2}) \kappa_{sc} v]}.\label{finalsol}
\end{eqnarray} 
The leading term in expansion of the above i.e. $\mathcal{O}(x^0)$ yields $f^{sc}_{lm}$.
Finally the complete solution of $F$ will be,
\begin{eqnarray}
 F(v,\theta,\phi) = \sum_{l,m} c^{(1)}_{kerr} \exp[(\Omega+ \frac{b(l,m) x^2}{2 \Omega}) \kappa_{sc} v] Y_{lm}(\theta,\phi).\nonumber\\
 \label{completeF} 
 \end{eqnarray}
One may check that the above one automatically makes $\delta\beta_A$ in (\ref{corr}), vanishing at the horizon for the choice (\ref{R}).
 Hence the equation (\ref{f}) shows that the dynamics of each Goldstone mode $f_{lm}$ is governed by inverted harmonics potential. Following argument described in the references \cite{Morita:2018sen}-\cite{Morita:2019bfr}, near horizon modes living in the inverted harmonic potential can be related to  thermal nature of the horizon through its chaotic dynamics. Therefore, one can identify the temperature perceived by every individual mode near the horizon of the black hole as,
\begin{eqnarray}
T(lm) = \frac{\hbar}{2 \pi} \Omega(l) \kappa_{sc}  \big( 1+ \frac{b(l,m) x^2}{2 \Omega^2(l)} \Big)~.\label{T1}
\end{eqnarray}
Detailed discussions can be found in Appendix G.
The average temperature perceived by individual $l$ mode can be calculated from the expression (\ref{T1}) as follows,
\begin{eqnarray}
&& T_{avg}=  \frac{\hbar}{2 \pi}\kappa_{sc} \Big( \frac{\sum_{l} \Omega(l)}{\sum_l 1} + \frac{x^2}{2 \sum_l (2l+1)} 
\nonumber
\\
&&\times \sum_l \sum_{m=-l}^{l} \frac{b(l,m)}{\Omega(l)}\Big)=\frac{\hbar}{2 \pi}\kappa_{sc} (\sqrt{\frac{3}{4}}- \frac{7}{25} x^2)
\nonumber
\\
&&= \frac{\hbar}{2 \pi}\kappa_{sc} \sqrt{\frac{3}{4}}(1- 0.32 x^2).\label{T}
\end{eqnarray}

To this end it would be illuminating to compare the above expression for the mode temperature due to underlying Goldstone mode degrees of freedom, with that of Hawking temperature of Kerr black hole \cite{Hawking:1974rv},
$
T_{BH} = {\hbar \kappa}/{2 \pi} .
$
For slowly rotating ($x =a/M \ll 1$)  Kerr black hole, using $ r_H/M \approx (2-\frac{x^2}{2})$, the surface gravity $\kappa$ is expanded as,
 \begin{eqnarray}
 \kappa&=& \frac{\sqrt{M^2-a^2}}{4 M r_H} \approx \frac{1-\frac{x^2}{2}}{2M (1-\frac{x^2}{4})} 
 \nonumber
 \\
& \approx& \kappa_{sc} \Big(1-\frac{1}{4} x^2 +\mathcal{O} (x^4) \Big). \label{kap}
\end{eqnarray}
  In this approximation the expression of the Hawking temperature for Kerr black hole can be expressed as the correction to the Hawking temperature of Schwarzschild black hole horizon upto $\mathcal{O}(x^2)$ as follows ,
\begin{eqnarray}
T_{BH} = \frac{\hbar}{2 \pi} \kappa_{sc} \Big(1-\frac{1}{4} x^2  \Big).\label{TH1}
\end{eqnarray}
Therefore comparing (\ref{T}) with (\ref{TH1}), the obtained expression of the temperature emerges out of unstable quantum dynamics of Goldstone modes are resemble with the usual Hawking temperature for slowly rotating black hole with different numerical constant, which is matching with our previous analysis.  
Our present analysis again suggests the fact that the thermal nature of the horizon of a black hole spacetime is intimately tied with the symmetry breaking and is associated with the emergence of super-translation Goldstone modes in the gravitational sector.  

\subsection{ZAMO observer trajectory}
In this section we study the dynamics of the Goldstone mode $F$ with respect to  the trajectory of zero angular momentum observers. In stationary rotating spacetime within the ergo sphere region, observers will not be static due to frame dragging effect, rather will be co-rotating with the black hole. In this section will consider the family of those co-rotating observers with zero angular momentum (ZAMO) in their proper frame. The trajectory of a ZAMO observer is defined by $r=$ constant and $\theta=$ constant $=\theta_c$ (say) (more detail can be found in \cite{Paddy}). In this scenario the aforementioned Goldstone mode $F$ is function of $v$ and $\phi$ only.  
For those observers, the Lagrangian (\ref{LF1}) reduces to,
\begin{eqnarray}
&&\mathcal{L}_{F} \vline_{\theta=\theta_c}= c_1(\theta_c) (\partial_v F)^2 + c_3(\theta_c) (\partial_{\phi} F)^2 \nonumber
\\
& +& c_5(\theta_c) (\partial_v \partial_{\phi} F)^2 +  c_7(\theta_c) \partial_{\phi} F\partial^2_v F + c_9(\theta_c) \partial^2_{\phi} F\partial^2_v F.\nonumber\\ 
\label{LF1ZAMO}
\end{eqnarray} 
Hence equation of motion will be,
\begin{eqnarray}
&&\Big(\chi^2_H(B (\theta_c) \partial^2_v F - 16 \Sigma_H^3 \partial^2_v \partial^2_{\phi} F) 
\nonumber
\\
&&+ 12 \alpha^2_2  \Sigma^3_H \partial^2_{\phi}F \Big)\vline_{\theta=\theta_c}=0~.
\label{ZAMOeqn}
\end{eqnarray}
 The expression of $B$ is given in Eq. (A3) of Appendix A. Using the azimuthal symmetry the solution can be chosen as $F (v,\phi)= \sum_{m} f_{m}(v)\exp(i m \phi)$, such that\ the equation of motion for $F$ given in (\ref{ZAMOeqn}) corresponds to a particular ZAMO observer situated at fixed angle $\theta_c$. Different value of $\theta_c$ yields different ZAMO observer. It is then reasonable to talk about an average equation over the all observers. It has been shown in Appendix H that after averaging over the different directions of $\theta$, the equation of motion for $f_{m}$ yields,
\begin{eqnarray}
\frac{\partial^2 f_m}{\partial v^2} -3   N(m) \kappa^2 f_m=0.\label{ZAMO12}
\end{eqnarray}
where $\kappa$ is the surface gravity of Kerr black hole. The expression of $N(m)$ is given in Appendix H.
 The equation (\ref{ZAMO12}) shows that the dynamics of $f_{m}$ is governed by inverse harmonic potential with solution,
\begin{eqnarray}
&&f_{m}(v)= \bar{A}_1 \exp{[\sqrt{3 N(m)}   \kappa v]}\nonumber
\\
&&+ \bar{A}_2 \exp{[-\sqrt{3 N(m)}   \kappa v]};\label{fsolZAMO}
\end{eqnarray}
Here $\bar{A}_1$ and $\bar{A}_2$ are two undetermined constant of integration having dimension of length.
Since we are interested to the near horizon region where $ v \rightarrow -\infty$, we have to discard second part of the solution (\ref{fsolZAMO}) which grows rapidly and makes the mode unstable. Therefore suitable boundary condition can be set as $\bar{A}_2=0$. 
So the complete solution of $F$  will be,
\begin{eqnarray}
F(v,\phi)= \sum_{m} \bar{A}_1 \exp{[\sqrt{3 N(m)}  \kappa v +i m \phi]}.\label{solZAMO}
\end{eqnarray}
Following the discussion in Appendix G about the connection between thermality and dynamics of chaotic system in the semi classical regime, the temperature in the present case is given by,
\begin{eqnarray}
T_{ZAMO}(m)= \frac{\hbar}{2 \pi}  \sqrt{3N(m)}  \kappa~.\label{TZAMO}
\end{eqnarray}
 Therefore, in the present discussion the thermal nature of the black hole horizon is captured through the quantum dynamics of the possible candidates of the BH microstates which are conjectured to be the Goldstone mode $F$ associated with the breaking of super-translation symmetry near the horizon. 


\section{Conclusion}
Due to spontaneous breakdown of global symmetry emergence of Goldstone modes, and their dynamics play fundamental role in many branches of physics. This same phenomena has recently gained widespread interest in the gravitational physics stimulated by the discovery of a beautiful connection between infinite dimensional BMS symmetry at null infinity and the well known soft graviton theorem \cite{Weinberg:1965nx}-\cite{Campiglia:2015qka}.  Due to spontaneous breaking of BMS symmetry, soft graviton modes have been conjectured to play as Goldstone modes in the black hole background. In our previous paper \cite{Maitra:2019eix} we used this very idea of spontaneous symmetry breaking near the black black hole horizon instead of asymptotic null infinity and investigated the dynamics of those mode. We considered spherically symmetric black hole spacetime. It turned out that in the free field limit, Goldstone mode dynamics is governed by the inverted harmonic potential. At the quantum level this instability \cite{Morita:2019bfr} is interpreted as the deep rooted cause of thermodynamic nature of the underlying black holes. Through our analysis, we could define an average effective thermodynamic temperature $T_{avg}$, which turned out to be $T_{avg} = \sqrt{3/4} ~T^{sc}_{BH}$, with $~T^{sc}_{BH}$ as the Bekenstein-Hawking temperature of the Schwarzschild black hole. 

In our present work, we extended the aforementioned analysis for rotating black hole. In the slow rotation limit, following the same methodology, we have arrived at perturbatively corrected temperature as,
\begin{eqnarray}
 T_{avg}= \sqrt{\frac{3}{4}}\left(1- 0.32 \left(\frac{a}{M}\right)^2\right) T^{sc}_{BH}.\label{TT}
\end{eqnarray} 
In the second part of this paper, we have analyzed for a special class of ZAMO observers and leads to thermalization of the BMS modes whose temperature is again proportional to Hawking expression,
\begin{eqnarray}
T_{ZAMO}= \frac{\hbar}{2 \pi}  \sqrt{3N(m)}  \kappa~ = \sqrt{3N(m)} T_{BH}^{kerr}.\label{TZAMO2}
\end{eqnarray} 
These are the two main results (Eqs. (\ref{TT}), (\ref{TZAMO2})) of our present analysis. Although our obtained results of the horizon temperature is not exactly matching with  Hawking expression, it is an interesting hint to investigate further on the symmetry breaking phenomena in the gravitational sector to understand the deeper underlying reasons of the thermodynamic nature of black holes. Our analysis seems to suggest in the direction where all the $(lm)$ modes of $F$ could be the underlying microstates responsible for the horizon entropy which requires in depth investigation. 

Also there are some important observations that should be mentioned here. In the present analysis we did a quantum mechanical treatment by considering the Schrodinger equation corresponding to the Goldstone mode $F$ to explore the thermal behaviour. However the parameter $F$ can be treated as quantum field which we leave for our further study. But it is expected that the behaviour of each mode of the quantum field is similar to the quantum mechanical wave function as far as temperature is concerned. Therefore the predicted thermalization and temperature in this analysis is expected to be well defined within the present analysis. Of course the present treatment will be complete once the field theoretical description will be done in which the definition of relevant vacuum state will be cleared. 
Another important comment can be mentioned here.
The  symmetry analysis has been performed for the near horizon Kerr metric constructed in GNC. It is obvious that only those observers sitting in this coordinate system will identify these symmetries and correspondinly this vector near the horizon. Thus the present symmetry analysis is  observer dependent. In this sense among all possible diffeomorphism symmetries only a subset has been chosen by our GNC observer which incorporates a thermalization of horizon at the semi-classical level and thereby is providing the observer dependence of the thermal nature of black hole.

\vskip 3mm
\noindent
{\bf Acknowledgements:}
The research of BRM is partially supported by a START- UP RESEARCH GRANT (No.
SG/PHY/P/BRM/01) from the Indian Institute of Technology Guwahati, India and by a Core Research Grant
(File no. CRG/2020/000616) from Science and Engineering Research Board (SERB), Department of Science $\&$
Technology (DST), Government of India. 




\newpage
 \begin{widetext}
\appendix
\begin{center}
{\bf{Supplementary material}}	
\end{center}



\section{Lagrangian (\ref{LF1})}\label{App3}
For the choice (\ref{R}) the Einstein-Hilbert Lagrangian  (\ref{L}) reduces to the following form after taking near horizon limit,
\begin{eqnarray}
&&\mathcal{L}_{(F,R^{\theta}=0,,R^{\phi}=C)}= c_1(\theta) (\partial_v F)^2 +c_2(\theta) (\partial_{\theta} F)^2 + c_3(\theta) (\partial_{\phi} F)^2 + c_4(\theta) (\partial_v \partial_{\theta} F)^2 +c_5(\theta) (\partial_v \partial_{\phi} F)^2 \nonumber
\\
&+& c_6(\theta) \partial_{\theta}F \partial^2_v F + c_7(\theta) \partial_{\phi} F\partial^2_v F +  c_8(\theta) \partial^2_{\theta}F \partial^2_v F + c_9(\theta) \partial^2_{\phi} F\partial^2_v F +  c_{10}(\theta) \partial_{v} F\partial^2_v F+ c_{11}(\theta) \partial_{\theta} F\partial_v \partial_{\theta} F \nonumber
\\
&+& c_{12}(\theta) \partial_{\phi} F\partial_v \partial_{\phi} F + c_{13}(\theta) \partial^2_{\theta} F \partial_v F +c_{14}(\theta)\partial_{\theta} F \partial_v F + c_{15}(\theta)  \partial^2_{\phi} F \partial_v F +c_{16}(\theta) \partial_{\phi} F \partial_v F .\label{LF}
\end{eqnarray}
$\mathcal{L}_{(F,R^{\theta}=0,,R^{\phi}=C)}$ is calculated on the stretched horizon at $\rho=$ constant surface, then the near horizon limit $\rho \rightarrow 0$ has be taken. Therefore action has been defined as the integration of the Lagrangian over the coordinates $v$, $\theta$ and $\phi$. The reduced form of the Lagrangian (\ref{LF}) correspond to the action for the super-translation mode $F$. In the expression (\ref{LF}) all the terms containing first derivative of $F$ with respect to $v$ can be expressed as total derivative. So last seven terms can be removed from the Lagrangian. Finally the  Lagrangian (\ref{LF}) turns into  (\ref{LF1}).
The expression $c_1....c_9$ are given by,

\begin{eqnarray}
	&&c_1=\frac{B}{2 \Sigma^2_H \chi_H \sin\theta} ;~~~~~ c_2=-6 \alpha^2_2 (\sin\theta) /(\chi_H \Sigma_H  );~~~~ c_3= -6\alpha_2^2 \Sigma_H  (\csc\theta) / \chi_H^3;~~~~~c_4= -6\chi_H (\sin\theta) /\Sigma_H;\nonumber
	\\
	&&c_5=-6 \Sigma_H  (\csc\theta) /\chi_H^2;~~~~~c_6=4  \chi_H (\cos\theta) /\Sigma_H ;~~~c_8=4 \chi_H (\sin\theta) / \Sigma_H;~~~~c_9=4 \Sigma_H  (\csc\theta) / \chi_H.
	\end{eqnarray} 
where,
\begin{eqnarray}
&& B= -\alpha_2 \sin^2\theta\Big( 4 r^2_H (2 r^3_H +a^2(M+3 r_H)) +a^4 (M+7 r_H)+4 a^2 r^2_H \alpha_2 \cos2\theta +a^4 \alpha_2 \cos4\theta \nonumber
\\
&& +8 r_H \chi_H (r^2_H+a^2 \cos2\theta)\Big);\nonumber\\ \label{B}
\\
&&\alpha_1= (M+r_H);~~\alpha_2= (r_H-M);~~ \alpha_3=(a^2+ M r_H)\nonumber
\end{eqnarray}


\section{GHY boundary term}\label{App4}
The GHY term is given by (\ref{GHY}).
For the corrected metric (\ref{corrected}), the lower components of $M^a$  are given by $ M_a =(0,1/\sqrt{2r(\kappa +\kappa \partial_v F +\partial^2_{v} F)},0,0)$.
Hence from GHY term we can write action in the following form by taking near horizon limit ($\rho \rightarrow 0$),
\begin{eqnarray}
\mathcal{S}_2= -\frac{\chi_H}{8 \pi G}\int d^3 x \sin \theta \Big[\kappa +\Big(\kappa \p_v F + \frac{1}{2}\p^2_v F + \frac{1}{2 \kappa} \p^3_v F\Big) + \frac{1}{2 \kappa
	^2}\Big(\kappa^2 \p_v F \p^2_v F + \kappa (\p^2_v F)^2 + \kappa \p_v F \p^3_v F 
+ \p^2_v F \p^3_v F  \Big)\Big].\nonumber\\
\label{Kac}
\end{eqnarray}
The first term is a constant, while the terms in the first bracket are the total derivative in $v$. Other terms can also be transformed into total derivative as follows,
\begin{eqnarray}
&&\p_v F \p^2_v F= \frac{1}{2} \p_v [(\p_v F)^2];~~~~\p^2_v F \p^3_v F= \frac{1}{2} \p_v [(\p^2_v F)^2]\nonumber
\\
&& (\p^2_v F)^2 + \p_v F \p^3_v F =\p_v [\p_v F \p^2_v F]. 
\end{eqnarray}
Hence all these total derivative terms should not contribute to the dynamics of $F$. However the boundary term added to the action could be related to the horizon entropy which is discussed in Appendix \ref{app1}. 


\section{\label{app1} Contribution of boundary term in the heat content of the horizon}

In the section \ref{lag} the GHY boundary term has been evaluated for our metric and added to the EH action which does not have effect on the mode dynamics. However this boundary term gains importance in the description of surface hamiltonian, defined as $H_{sur} = -\frac{\partial S_2}{\partial v}$, which is directly related to the heat content of the horizon \cite{Padmanabhan:2012bs}-\cite{Majhi:2013jpk}. Hence to show the aforementioned connection with the heat content in the present analysis, we have to calculate the form of the Hamiltonian for the mode solution of $F$ given in (\ref{completeF}). So putting this form of $F$ in the GHY action $S_2$ (\ref{Kac}), after integrating we get the resultant expression of the Hamiltonian as the following,
 \begin{eqnarray}
H_{sur}=\frac{\bar{A}_{kerr}}{8 \pi G} \Big[ \kappa + f_3(\kappa,x^2) 
 e^{f_2(\kappa, x^2) v}\Big]~,\label{H}
\end{eqnarray}
 $\bar{A}_{kerr}$ denotes the transverse area of the horizon of kerr spacetime.
In the evaluation of the Hamiltonian $H_{sur}$, we found out that all the first and higher derivative terms of $F$  will generate some complicated expressions (functions of $\kappa$ and $x^2$) which are exponentials in $v$. However the explicit forms of these terms are not needed in our present discussion as all these terms are exponentially suppressed towards the horizon in the limit $v\rightarrow -\infty$. Therefore to show the nature of those terms in the Hamiltonian, we have kept the form $f_3(\kappa,x^2) e^{f_2(\kappa, x^2) v}$ in (\ref{H}), where the functional form of $f_3(\kappa,x^2)$ and $f_2(\kappa, x^2)$ are not explicitly given but they are finite near the horizon. Hence the second term in (\ref{H}) vanishes near the boundary. The above Hamiltonian then is simplified into,
\begin{eqnarray}
 H_{sur} = \frac{1}{8 \pi G} \bar{A}_{kerr} \kappa~. \label{TS}
\end{eqnarray} 
It is known that the  horizon entropy and the temperature of the horizon are given by $S = \bar{A}_{kerr} /4 G $  and $T = \kappa/2\pi$ respectively. Hence with the help of these two expressions, we can write the surface Hamiltonian  (\ref{TS}) as,
\begin{eqnarray}
H_{sur}= TS.
\end{eqnarray} 
This result clearly shows that the GHY boundary term in the action is directly connected with the heat content of the horizon.


\section{Details of Eq. (\ref{kerrF})} \label{feq}

After rearranging the terms in the equation (\ref{kerrF}) and simplifying, we can write,
\begin{eqnarray}
 &&  \Big[-16  \chi^4_H (\chi^2_H -2a^2 \chi_H \sin^2\theta +a^4 \sin^4\theta ) \p^2_{\theta} \p^2_v F -16 \cot\theta \chi^4_H (\chi^2_H \nonumber
 \\
 &&-a^4 \sin^4\theta ) \p_{\theta} \p^2_v F-\frac{16 \chi^2_H}{\sin^2\theta}( \chi^4_H -4 a^2 \chi^3_H \sin^2\theta +6 a^4 \chi^2_H \sin^4\theta-4 a^6 \chi_H \sin^6\theta  \nonumber
 \\
 &&+a^8  \sin^8\theta ) \p^2_{\phi} \p^2_v F  +\alpha\chi^2_H \p^2_v F \Big]\nonumber
 \\
 && + 12 \alpha^2_2   \Big[  \chi^2_H (\chi^2_H -2a^2 \chi_H \sin^2\theta+a^4 \sin^4\theta) \p^2_{\theta} F  +\cot\theta \chi^2_H (\chi^2_H \nonumber
 \\
 &&-a^4 \sin^4\theta) \p_{\theta} F + \frac{1}{\sin^2\theta}( \chi^4_H -4 a^2 \chi^3_H \sin^2\theta +6 a^4 \chi^2_H  \sin^4\theta\nonumber
 \\
 &&-4 a^6 \chi_H \sin^6\theta  +a^8  \sin^8\theta ) \p^2_{\phi} F  \Big]=0\label{eqker}
\end{eqnarray}
In the above equation we have used the expansion of the quantities as, 
\begin{eqnarray}
&& \Sigma^2_H = (\chi_H -a^2 \sin^2\theta)^2=\chi^2_H -2a^2 \chi_H \sin^2\theta +a^4 \sin^4\theta.\nonumber
\\
&&\Sigma^4_H = (\chi^2_H-a^2 \sin^2\theta)^4= \chi^4_H -4 a^2 \chi^3_H \sin^2\theta +6 a^4 \chi^2_H \sin^4\theta \nonumber
\\
&& -4 a^6 \chi_H \sin^6\theta  +a^8  \sin^8\theta. 
\end{eqnarray} 
Now the equation (\ref{eqker}) can be more simplified and written as,
\begin{eqnarray}
&& \Big[ -16\chi^6_H \Big( \p^2_{\theta}\p^2_v F +\cot\theta \p_{\theta} \p^2_v F +\frac{1}{\sin^2\theta} \p^2_{\phi} \p^2_v F \Big) -16 \chi^4_H (-2a^2\chi_H \sin^2\theta\nonumber
\\
&& +a^4\sin^4\theta) \p^2_{\theta} \p^2_v F +(16a^4 \chi^4_H \sin^4\theta \cot\theta) \p_{\theta} \p^2_v F - \chi^2_H  (-64  a^2 \chi^3_H \nonumber
\\
&&+ 96  a^4 \chi^2_H \sin^2\theta-64  a^6 \chi_H \sin^4\theta + 16 a^8  \sin^6\theta +\alpha) \p^2_v \p^2_{\phi} F \Big]\nonumber
\\
&& + 12 \alpha^2_2 \Big[ \chi^4_H \Big( \p^2_{\theta} F +\cot\theta \p_{\theta} F +\frac{1}{\sin^2\theta} \p^2_{\phi} F \Big) + \chi_H^2 (-2a^2 \chi_H \sin^2\theta\nonumber
\\
&& + a^4 \sin^4\theta) \p^2_{\theta} F - (a^4 \chi^2_H \sin^4\theta \cot\theta) \p_{\theta} F -(4 a^2 \chi^3_H \nonumber
\\
&& -6 a^4 \chi^2_H \sin^2\theta + 4 a^6 \chi_H \sin^4\theta - a^8  \sin^6\theta ) \p^2_{\phi} F \Big]=0.\label{eqnk2}
\end{eqnarray}  
From the above equation one can derive dynamical equation of $F$ in slow rotation approximation as shown in the Appendix \ref{slow0}.

\section{Results in slowly rotating background}
\subsection{Derivation  of the Eq.  (\ref{slow2}) from Eq. (\ref{kerrF})  }\label{slow0}
In slowly rotating background for  $a \ll M$, we have considered $a/M \approx x$, which imples that 
\begin{eqnarray}
\frac{r_H}{M} =1 + \sqrt{1-\frac{a^2}{M^2}} =  1 + \sqrt{1-x^2} \approx 1+1-\frac{1}{2} x^2= 2-\frac{1}{2} x^2.\label{rhm}
\end{eqnarray}
Now using (\ref{rhm}), from the equation (\ref{eqnk2}) we expand the term $\chi^6_H$ and also the coefficient of $\p_{\theta}^2 \p^2_v F$ respectively in slowly rotation approximation as follows,
\begin{eqnarray}
\chi_H^6= (a^2+r^2_H)^6= 2^6 M^6 r^6_H= 2^6 M^{12} (\frac{r_H}{M})^6 =2^6 M^{12} (2-\frac{1}{2} x^2)^6= 2^{11} (2-3 x^2)+ \mathcal{O} (x^4).
\end{eqnarray}
Also,
\begin{eqnarray}
&&  2 a^2 (a^2+r^2_H)^5 \sin^2\theta -a^4 (a^2+r^2_H)^4 \sin^4\theta = 2^6 a^2 M^5 r^5_H \sin^2\theta \nonumber
\\
&&- (2aM r_H \sin\theta)^4 =2^6 M^{12} (\frac{a}{M})^2 (\frac{r_H}{M})^5 \sin^2\theta -2^4 M^{12} (\frac{a}{M})^4 (\frac{r_H}{M})^4 \sin^4\theta\nonumber
\\
&& = 2^6 M^{12} x^2 (2-\frac{1}{2} x^2)^5 \sin^2\theta- 2^4 M^{12} x^4 (2-\frac{1}{2} x^2)^4 \sin^4\theta\nonumber
\\
&& \approx  2^{11} M^{12} x^2 \sin^2\theta +\mathcal{O} (x^4).\nonumber\\ \label{slow1}
\end{eqnarray}
Similarly the other terms of (\ref{eqnk2}) can be expanded upto  $\mathcal{O} (x^2)$  keeping only $M$ and $x$ in the equation. Thus from (\ref{eqnk2}) we get the equation (\ref{slow2}).



\subsection{Derivation of Eq. (\ref{Y0})}\label{App6}
Eq. (\ref{slowly}) for all the modes $l \geq 1$, can be expressed in the following form using the various well known properties of spherical harmonic functions. The derivative of spherical harmonics can be written as,

	\begin{eqnarray}
\sin^2\theta (\p^2_{\theta} Y_{lm}) &=& (m^2 \cos^2\theta-m) Y_{lm} +\sqrt{(l-m)(l+m+1)} (2m+1) e^{i \phi}\sin\theta \cos\theta~ Y_{l(m+1)} +\nonumber
	\\
	&& \sqrt{(l-m)(l-m-1)(l+m+2)(l+m+1)} \sin^2\theta e^{2 i \phi}~ Y_{l(m+2)}.\label{derivatve}
	\end{eqnarray}
Now we can express $\cos^2\theta$ and $ \sin\theta \cos\theta$ in terms of spherical harmonics function as,
\begin{eqnarray}
\sin^2\theta \exp[2 i \phi]= \frac{4 \sqrt{2 \pi}}{\sqrt{15}} Y_{22}\nonumber\\
\sin\theta \cos\theta \exp[{i \phi}]  = -\frac{2 \sqrt{2\pi}}{\sqrt{15}} Y_{21}\nonumber\\
\cos^2\theta= \frac{1}{2\sqrt{\pi}}Y_{00} + \frac{1}{\sqrt{5\pi}} Y_{20}.\label{harmonics}
\end{eqnarray}
Where the use has been made of the following contraction rule of spherical harmonics,
\begin{eqnarray}
Y_{l_1m_1} Y_{l_2m_2}= \sum_{l_3m_3}\Lambda^{l_1 m_1}_{l_2 m_2 l_3 m_3} Y_{l_3m_3} ; \label{3j}
\end{eqnarray}
where $\Lambda$ is expressed by the Wigner 3-j symbols defined for the product of two spherical harmonics as,

	\begin{eqnarray}
	\Lambda^{l_1 m_1}_{l_2 m_2 l_3 m_3} = \sqrt{\frac{(2l_1+1)(2l_2+1)(2l_3+1)}{4\pi}}  
	\begin{pmatrix}
	l_1 &l_2 & l_3\\
	m_1 &m_2 &m_3
	\end{pmatrix}
	\begin{pmatrix}
	l_1 &l_2 & l_3\\
	0&0&0
	\end{pmatrix}.
	\end{eqnarray}
The selection rules on Wigner 3-j symbols are given by $\mid l_1- l_2 \mid \leq l_3 \leq \mid l_1+ l_ 2 \mid$ and $ m_3=m_1+m_2$.

Now to get the equation of motion for generic mode $f_{lm}$,  we use the above relations between trigonometric functions and spherical harmonics and also the result found from  the second order derivative of $Y_{lm}$ with respect to $\theta$ as given (\ref{derivatve}). Hence by substituting (\ref{derivatve}) and (\ref{harmonics}) in (\ref{slowly}) and considering constant and $x$ dependent parts of the equation (\ref{slowly}) separately, the equation can be expressed as,
\begin{eqnarray}
&& \sum_{lm}\Big[ 32 (l(l+1) -1 ) \p^2_v f_{lm} -\frac{3}{M^2} l(l+1) f_{lm}\Big] Y_{lm} \nonumber
\\
&& + x^2 \Big[\p^2_v f_{lm} \Big(-96 l(l+1) Y_{lm} + 8 \pi  (\frac{1}{2\sqrt{\pi}} Y_{00} +\frac{1}{\sqrt{5\pi}} Y_{20}) Y_{lm} \nonumber
\\
&& +\frac{128\pi}{3} \Big(m^2  (\frac{1}{2\sqrt{\pi}} Y_{00} +\frac{1}{\sqrt{5\pi}} Y_{20}) -m \Big)Y_{lm} \nonumber
\\
&& +64\sqrt{\frac{2\pi}{15}} \sqrt{(l-m)(l+m+1)} (2m+1)  \times Y_{2-1} Y_{l~m+1} \nonumber
\\
&& +128  \sqrt{\frac{2\pi}{15}} \sqrt{(l-m)(l-m-1)(l+m+2)(l+m+1)} Y_{2-2} Y_{l~m+2}\nonumber
\\
&& -2(32 m^2  -37) Y_{lm} \Big) -\frac{1}{M^2} f_{lm} \Big(-6 l(l+1) Y_{lm}
\nonumber
\\
&& +2\pi \Big(m^2  (\frac{1}{2\sqrt{\pi}} Y_{00} +\frac{1}{\sqrt{5\pi}} Y_{20})-m \Big) Y_{lm} \nonumber
\\
&&+3 \sqrt{\frac{2\pi}{15}} \sqrt{(l-m)(l+m+1)} (2m+1) Y_{2-1} Y_{l~m+1}  +6 \sqrt{\frac{2\pi}{15}} \nonumber
\\
&& \times \sqrt{(l-m)(l-m-1)(l+m+2)(l+m+1)} Y_{2-2} Y_{l~m+2} -3 m^2 Y_{lm} \Big)\Big]=0.\nonumber\\
\label{slow4}
\end{eqnarray}
Now we use (\ref{3j}) to calculate $Y_{20} Y_{lm}$,~ $Y_{2-1} Y_{l~m+1}$ and $Y_{2-2} Y_{l~m+2}$~ as follows, 
\begin{eqnarray}
&&Y_{20} Y_{lm} =\sum _{m_3=-l_3}^{l_3} \sum_{l_3=\mid 2- l \mid}^{\mid 2+ l \mid}\Lambda^{2 0}_{lm l_3 m_3} Y_{l_3m_3}; \nonumber
\\
&& Y_{2-1} Y_{lm+1} =\sum _{m_3=-l_3}^{l_3} \sum_{l_3=\mid 2- l \mid}^{\mid 2+ l \mid}\Lambda^{2 -1}_{lm+1 l_3 m_3} Y_{l_3m_3};\nonumber
\\
&& Y_{2-2} Y_{lm+2} =\sum _{m_3=-l_3}^{l_3} \sum_{l_3=\mid 2- l \mid}^{\mid 2+ l \mid}\Lambda^{2 -2}_{lm+2 l_3 m_3} Y_{l_3m_3}.\label{comy}
\end{eqnarray}
 Then using the above relations we express the  equation (\ref{slow4}) in the combined spherical harmonic basis as follows,
\begin{eqnarray}
&&\sum_{lm}\Big[  (l(l+1) -1) \p^2_v f_{lm} -\frac{3}{32 M^2} l(l+1) f_{lm}\Big] Y_{lm}  \nonumber
\\
&&  +\sum_{lm} x^2 \Big[\beta_1(l,m) \p^2_v f_{lm} +\beta_2 (l,m) f_{lm} + \sum_{m_3=-l_3}^{l_3} \sum_{l_3= \mid l -2 \mid} ^{\mid l + 2 \mid}  \Big( \beta_{lml_3m_3}(l,m) \p^2_v f_{lm}\nonumber
\\
&& +\bar{\beta}_{lml_3m_3} (l,m) f_{lm} \Big) \Big]Y_{l_3m_3}=0.\label{slow5}
\end{eqnarray}
where we have defined following quantities,
\begin{eqnarray}
	&& \beta_1 (l,m)=  \Big( \frac{-72l(l+1) +57-40 m^2-24 m}{24}  \Big);\nonumber
	\\
	&& \beta_2(l,m)= \Big( \frac{12 l(l+1) +5 m^2+3 m}{64 M^2} \Big);\nonumber
	\\
	\nonumber\\
	&&\beta_{lml_3m_3} (l,m)= 2\sqrt{\pi} [ \Lambda^{20}_{lml_3m_3} (\frac{2 m^2+24}{3\sqrt{5}}) +(2m+1) \Big( \Lambda^{2-1}_{l(m+1)l_3m_3}\nonumber
	\\
	&& \times (\sqrt{\frac{2(l-m)(l+m+1)}{15}} ) \nonumber
	\\
	&& +2 \Lambda^{2-2}_{l(m+2)l_3m_3}  \sqrt{\frac{2(l^2-(m+1)^2) (l(l+2)-m(m+2))}{15}} \Big)]; \nonumber
	\\
	\nonumber\\
	&&\bar{\beta}_{lml_3m_3} (l,m)= 2\sqrt{\pi} [ \Lambda^{20}_{lml_3m_3} \frac{ m^2}{32 \sqrt{5} M^2} +\Lambda^{2-1}_{l(m+1)l_3m_3} \nonumber
	\\
	&& \times \frac{ 3(2m+1)\sqrt{2(l-m)(l+m+1)}}{64M^2\sqrt{15} } \nonumber
	\\
	&& + \Lambda^{2-2}_{l(m+2)l_3m_3}\frac{3(2m+1)\sqrt{2(l^2-(m+1)^2) (l(l+2)-m(m+2))}}{32 M^2\sqrt{15}}]. \nonumber\\ \label{Y11} 
	\end{eqnarray} 
Now  in the third and fourth terms in the second line of (\ref{slow5}), at first we replace dummy indices as $ l \rightarrow l'$ and $ m \rightarrow m'$ , after that we again replace $l_3 \rightarrow l$ and $m_3 \rightarrow m$ . Thus finally from (\ref{slow5}) we get,
\begin{eqnarray}
&&\Big(\sum_{lm}\Big[  (l(l+1) -1) \p^2_v f_{lm} -\frac{3}{32 M^2} l(l+1) f_{lm}\Big]  \nonumber
\\
&&  +\sum_{lm} x^2 \Big[\beta_1(l,m) \p^2_v f_{lm} +\beta_2 (l,m) f_{lm}\Big] +x^2 \Big[\sum_{l'm'} \sum_{m=-l}^{l} \sum_{l= \mid l' -2 \mid} ^{\mid l' + 2 \mid}  \Big( \beta_{l'm'lm}(l',m') \p^2_v f_{l'm'}\nonumber
\\
&& +\bar{\beta}_{l'm'lm}(l',m') f_{l'm'} \Big) \Big] \Big)Y_{lm}=0.\label{slow6}
\end{eqnarray}
Now using the selection rule, one can write, 
\begin{eqnarray}
\sum_{m'=-l'}^{l'} \sum_{l'=0}^{\infty} \sum_{m=-l}^{l}  \sum_{l= \mid l' -2 \mid} ^{\mid l' + 2 \mid} \equiv  \sum_{m'=-l'}^{l'}  \sum_{l'= \mid l -2 \mid} ^{\mid l + 2 \mid} \sum_{m=-l}^{l} \sum_{l=0}^{\infty}.\label{sele} 
\end{eqnarray}  
Then using the above rule, the equation (\ref{slow6}) boils down to the equation (\ref{Y0}).


\section{Finding Eq. (\ref{f})}\label{App7}
Substituting (\ref{perturbsol}) in (\ref{Y1}) till $\mathcal{O}(x^2)$, we found that the zeroth order part of the equation which matches with the  equation of $f^{sc}_{lm}$ automatically vanishes. Then  the coefficients of $x^2$ in the equation (\ref{Y1}) gives the following equations,
\begin{itemize}
\item For $l=1$,
\begin{eqnarray}
	\p^2_{v} f^{(1)}_{1m} -\Omega^2 \kappa^2_{sc}  f^{(1)}_{1m} = c^{(0)}_{sc} \kappa_{sc} \Big[  \bar{b} \exp[\bar{\Omega}  \kappa_{sc} v] + \bar{b}_1 \exp[\bar{\Omega}_1  \kappa_{sc} v]  \Big],\label{perturb11}
	\end{eqnarray}
	where $-1 \leq m \leq 1$.
\item 	For  $l \geq 2$,\\
 the coefficients of $x^2$ in the equation (\ref{Y1}) reduces to the following form;	
\begin{eqnarray}
	\p^2_{v} f^{(1)}_{lm} -\Omega^2 \kappa^2_{sc}  f^{(1)}_{lm} = c^{(0)}_{sc} \kappa_{sc} \Big(b (l,m) \exp[\Omega  \kappa_{sc} v] + b_1(l,m) \exp[\Omega_1  \kappa_{sc} v] + b_2 (l,m) \exp[\Omega_2  \kappa_{sc} v] \Big).\label{perturb1}
	\end{eqnarray}
\end{itemize}
This is the equation of the first order perturbation $f^{(1)}_{lm}$.  The corresponding quantities in the above equations are expressed as follows,
	\begin{eqnarray}
	&&\Omega_1(l)= \Omega(l \rightarrow l+2);~~~~~\Omega_2(l)= \Omega(l \rightarrow \mid l- 2 \mid),~~~\bar{\Omega}=\Omega \vline_{l=1},~~~ \bar{\Omega}_1= \Omega_1 \vline_{l=1};\nonumber\\
	&&b (l,m)=  -\frac{1}{2l(l+1)-1} (\frac{-72l(l+1) +57-40 m^2-24 m}{24}\Omega^2(l) +\frac{12 l(l+1) +5 m^2+3 m}{4} )+ \Lambda^{20}_{lmlm}~ b_0(l,m) \nonumber
	\\
	&& + \Lambda^{2-1}_{l(m+1)lm}~ b'(l,m)+ \Lambda^{2-2}_{l(m+2)lm}~  b''(l,m);\nonumber
	\\
	&& b_0(l,m)=-\frac{2 \sqrt{\pi}}{\sqrt{5}(2l(l+1)-1)} (\frac{\Omega^2}{16}-m^2);~~~~ b'(l,m)= -\frac{2 \sqrt{\pi}}{(2l(l+1)-1)}(2m+1)\sqrt{\frac{2(l-m)(l+m+1)}{15}} \Big(\Omega^2-\frac{3}{4} \big);\nonumber
\\ 
	&& b''(l,m)= -\frac{2 \sqrt{\pi}}{2l(l+1)-1}  \sqrt{\frac{2(l^2-(m+1)^2) (l(l+2)-m(m+2))}{15}}\Big( 2 \Omega^2-\frac{3}{2}\Big);\nonumber
	\\
	&& b_1(l,m)= \Lambda^{20}_{(l+2)mlm}~ b_0(l\rightarrow l+2) + \Lambda^{2-1}_{(l+2)(m+1)lm}~ b'(l\rightarrow l+2)+ \Lambda^{2-2}_{(l+2)(m+2)lm}~  b''(l \rightarrow l+2);\nonumber
	\\
	&& b_2(l,m)= \Lambda^{20}_{(\mid l-2 \mid)mlm}~ b_0(l\rightarrow \mid l-2 \mid) + \Lambda^{2-1}_{(\mid l-2 \mid)(m+1)lm}~ b'(l\rightarrow \mid l-2 \mid)+ \Lambda^{2-2}_{(\mid l-2 \mid)(m+2)lm}~  b''(l\rightarrow \mid l-2 \mid);\nonumber
	\\
	&& \bar{b}= b\vline_{l=1},~~~ \bar{b}_1=b_1\vline_{l=1}.\nonumber	
	\end{eqnarray}


\begin{table}
\caption{Numerical values of the terms in the R.H.S of (\ref{perturb1}) and (\ref{perturb11}) and $\frac{b}{2\Omega}$} \label{table}
\begin{center}
\begin{tabular}{|c|c|c|c|c|c|}
	\hline
 l & m & b $\exp{[\Omega]}$ & $b_1 \exp{[\Omega_{1}]}$ & $b_2 \exp{[\Omega_{2}]}$ & $\frac{b}{2\Omega}$ \\ 
 \hline
 1& 0& -2.3 & 0.2 & | & -0.43\\
 1 & 1 & -1.6 & 0.1 &| & -0.5 \\
 1 &-1 & -2.8 & 0.2 &| & -0.26\\
 2 & 0 & -1.5 & 0.2  & -0.05 & -0.3\\
 2 & 1 & -1.3 & 0.15 & 0 & -0.32 \\
 2 & -1 & -1.5 & 0.18 & 0 & -0.3  \\
 2 & -2 & -1.16 & 0.1 & 0 & -0.35\\
 2 & 2 & -1.65 & 0.2 & 0 & -0.4\\
 $\vdots$ &  $\vdots$ & $\vdots$ & $\vdots$ & $\vdots$ & $\vdots$\\
 10 & 0 & -1.2 & 0.16 & 0.13 & -0.32\\
 10 & 10 & -1.13 & 0.04 & 0 & -0.3\\
 10 & -10 & -1.24 & 0.04 & 0 & -0.3\\
  $\vdots$ &  $\vdots$ & $\vdots$ & $\vdots$ & $\vdots$ & $\vdots$\\
  100 & 0 & -1.2 & 0.15 & 0.15 & -0.3\\
 100 & 100 & -1.2 & 0.005 & 0 & -0.3\\
 100 & -100 & -1.2 & 0.006 & 0 & -0.3\\
 \hline
\end{tabular}\\
\end{center}
\end{table}
It is clear that all the source terms which are present in the right hand side of the equations (\ref{perturb1}) and (\ref{perturb11}) are exponentially decaying and vanishes near the horizon in the limit $v \rightarrow -\infty$. However if we closely look at the numerical values of $b$, $b_1$ and $b_2$, it can be shown that $\mid b_1 \mid <\mid b \mid$ and also $\mid b_2 \mid < \mid b \mid$ for all ($l,m$). Therefore, the terms containing $b_1$ and $b_2$ are decaying faster than those terms containing $b$ as one approaches near the horizon. We have tabulated the numerical values of these source terms for few sample $(l,m)$ modes in the table \ref{table} with a particular choice that $\kappa_{sc} v=1$. It is observed that the numerical values of $b_1 \exp{[\Omega_{1}]}$ and $b_2 \exp{[\Omega_{2}]}$ are smaller compared to that of $b \exp{[\Omega]}$ in the near horizon limit. Therefore, near the horizon dominating contribution comes from the term $b \exp{[\Omega]}$ in (\ref{perturb1}) and also in (\ref{perturb11}). Hence neglecting other source terms, the equation of motion for $ f^{(1)}_{lm}$ boils down to ($l \geq 1$),
\begin{eqnarray}
\p^2_{v} f^{(1)}_{lm} -\Omega^2 \kappa^2_{sc}  f^{(1)}_{lm} = c^{(0)}_{sc} \kappa_{sc} b(l,m) \exp[\Omega  \kappa_{sc} v]= \kappa^2_{sc} b(l,m) f^{sc}_{lm}.\label{flmEqn}
\end{eqnarray}	
 Upto $\mathcal{O}(x^2)$, one can approximate, $x^2 f^{sc}_{km} \approx x^2 f_{lm}.$ Then multiplying the equation (\ref{flmEqn}) with $x^2$ and the adding  zeroth order equation of $f^{sc}_{lm}$ with it, the above equation can be written as the equation of motion of the complete $f_{lm}$ given in (\ref{f}).

	


\section{Thermal behaviour of the mode solution}\label{App8}
Now the question is whether the emergence of instability near the horizon of the rotating spacetime can be shown to be related with the thermal nature of the horizon. 
This analysis has been inspired by some recent conjecture \cite{Morita:2018sen} where the connection between semi-classical chaotic system and thermal behaviour has been discussed. For a chaotic system in attached with a thermal bath, characteristic Lyapunov exponent $\lambda$ is bounded from below as $\lambda \leq 2\pi T/\hbar$ \cite{Maldacena:2015waa}. By reversing this argument, quantum behavior of a classically chaotic system has been conjectured to be intrinsically thermal in nature with the temperature bounded by the Lyapunov exponent \cite{Morita:2019bfr}\cite{Kurchan:2016nju}. For this, modes living in the inverted harmonic potential have been shown to have transition probability obeying Bolzmann distribution with an effective temperature related to the strength of the inverted harmonic potential.

For our paper to be self contained with \cite{Maitra:2019eix} we describe here the motion of a classical particle in one-dimensional inverted harmonic potential. The equation of motion of the particle is given by;
\begin{eqnarray}
\mu \ddot{y} -\bar{\omega} y =0; \label{1dParticle}
\end{eqnarray}
here potential $V= -\frac{\bar{\omega} y^2}{2}$ and $\mu$ is the mass of the particle. Now the solution of the given equation of motion is;
\begin{eqnarray}
y(t) = a_1 e^{\sqrt{\bar{\omega}/\mu} t}+a_2 e^{-\sqrt{\bar{\omega}/\mu}t};
\end{eqnarray}
here $a_1$ and $a_2$ are arbitrary constants. Now we concentrate on the total energy $E$ of the particle moving under this potential. If $E > 0$, then classically the particle motion will be similar to that of a free particle motion with no effect of the potential on its trajectory. but if  $E<0$ then the potential energy of the particle is greater than its kinetic energy and the particle has to face the potential barrier. In that condition if the particle is moving toward the potential from left ($y <0$), classically it cannot pass through the potential towards right $(y >0$). But if we consider quantum mechanical treatment, then the amplitudes of the wave associated with the particle will be non zero in the other side of the barrier. So the particle will have finite probability of transmission through the barrier even for $E<0$. Similarly for $E>0$ particle will have finite probability of reflection off the barrier quantum mechanically.

So following the last discussion to describe the above quantum mechanical phenomena we can write the Schr\"{o}dinger equation for the wave function $\Phi(y)$ associated with the particle,
\begin{eqnarray}
-\frac{{\hbar}^2}{2}\frac{\partial^2 \Phi}{\partial y^2} - \frac{\bar{\omega}^2 y^2}{2} \Phi =E \Phi~. 
\end{eqnarray}
Hence the transmission coefficient ($T$) and the reflection cofficient ($R$) can be easily derived using WKB approximation (detail in \cite{Barton:1984ey}). Corresponding probabilities are given by , 
\begin{eqnarray}
P_{T/R} = \frac{1}{e^{\frac{2 \pi}{\hbar} \sqrt{\frac{\mu}{ \bar{\omega}}} |E|} +1} = \frac{1}{ e^{\beta |E|} +1}~. 
\end{eqnarray} 
For large absolute value of  $E$, probability amplitude from classical trajectories to quantum transmission or reflection will be $\exp[-\beta |E|]$. This probability amplitude  (Bolzmann factor) can be easily  attributed to a two level system with temperature $T$, whose ground state is represented as the classical path and excited state corresponds to the quantum one. The temperature of this two-level system can be easily defined by, 
\begin{eqnarray}
T = \frac{\hbar}{2 \pi} \frac{\bar{\omega}}{\mu} ~.
\end{eqnarray}

In our present analysis we have obtained the dynamical equation of motion for individual mode given in (\ref{f}). Comparing that equation with Eq. (\ref{1dParticle}) one can easily conclude that the dynamics of the mode is governed by inverted harmonic oscillator potential with the following identification:
\begin{eqnarray}
f_{lm} \equiv y; \,\,\,\,\ v \equiv t;
\end{eqnarray}
and also,
\begin{eqnarray}
\bar{\omega}^2 \equiv (\Omega^2+ b x^2) \kappa^2_{sc}~;~~\mu=1.
\end{eqnarray}
The mode dependent inverted harmonic potential takes the following form form,
\begin{eqnarray}
V_{lm}=-\frac{1}{2}  (\Omega^2+ b x^2) \kappa^2_{sc} f^2_{lm}.
\end{eqnarray}
Considering similar approach as described before, we can identify the temperature of our system given in Eq.(\ref{T1}).



\section{Derivation of Eq. (\ref{ZAMO12})}\label{App9}

After the substitution of the ansatz for $F$ in (\ref{ZAMOeqn}), for each mode $m$ the equation can be written as,
\begin{eqnarray}
& [B(\theta_c) +16 m^2 (r^2_H+a^2 \cos^2\theta_c)] \frac{\partial^2 f_{m}}{\partial v^2}-\frac{12 (M-r_H)^2 m^2}{(a^2+r^2_H)^2}  (r_H^2 +a^2 \cos^2 \theta_{c})^3 f_{m}(v)=0,\label{eqnZAMO1}
\end{eqnarray}
Hence the one dimensional equation of motion presented in (\ref{eqnZAMO1}) gives the dynamics of $f_m(v)$ for a specific ZAMO observer for a given angle $\theta_c$. However there will be a large number of ZAMO observers situated at different angles in the spacetime. Hence to study the mode dynamics of $f_m(v)$ corresponding to all those observers it is customary to take average over the angle $\theta$. We compute the required average by multiplying the equation by $\sqrt{g_{\theta\theta}}\vline_{\theta=\theta_c}=\sqrt{\Sigma_H (\theta_c)}$ and integrating  $\theta_c$ for all values ranging from $0$ to $\pi$. 
So from (\ref{eqnZAMO1}) we get,
\begin{eqnarray}
&&\frac{1}{\int_0^{\pi} \sqrt{\Sigma_H (\theta_c)} d\theta_c} \Big( \frac{\partial^2 f_m}{\partial v^2} \int_0^{\pi} [B(\theta_c) +16 m^2 (r^2_H+a^2 \cos^2\theta_c)] \sqrt{\Sigma_H (\theta_c)} d\theta_c\nonumber
\\
&&- f_m (v) \int_0^{\pi}  \frac{12 (M-r_H)^2 m^2}{(a^2+r^2_H)^2}  (r_H^2 +a^2 \cos^2 \theta_{c})^3 \sqrt{\Sigma_H(\theta_{c})} d\theta_c \Big) =0. \nonumber\\ \label{ZAMO11}
\end{eqnarray}
Then after averaging over all possible  ZAMO observers finally the equation of motion comes out as given in (\ref{ZAMO12}) where we have defined,
\begin{eqnarray}
N= m^2(P_1~ \textrm{Elliptic}E[-a^2/r_H^2] +P_2~  \textrm{Elliptic}K[-a^2/r_H^2])/ (P_3~ \textrm{Elliptic}E[-a^2/r_H^2] + P_4~ \textrm{Elliptic}K[-a^2/r_H^2])
\end{eqnarray}
where,
\begin{eqnarray}
&&P_1=2(-96 a^8 m^2 -2 r^6_H \alpha_2 (3M-17 r_H) +2a^2 r^4_H (M^2 - 135 M r_H + 2 r_H^2 (67 - 88 m^2) )\nonumber
\\
&&-2 a^4 r_H^2 (64 M^2 + 5 M r_H + 3 r_H^2(-23 + 88 m^2) ) -a^6 (69 M^2 - 110 M r_H + r_H^2 (41 + 368 m^2 ));\nonumber
\\
\nonumber\\
&&P_2= \chi_H (-96 a^6 m^2 + 4 r_H^4 (3 M^2 - 20 M r_H + 17 r_H^2) -2 a^2 r_H^2 (M^2 + 40 M r_H + r_H^2(-41 + 142 m^2) )\nonumber
\\
&&-a^4 (69 M^2 - 110 M r_H + r_H^2 (41 + 284 m^2) ));\nonumber
\\
\nonumber\\
&&P_3= 8 a^6(6 a^2 + 23 r_H^2) + 11 a^2 r_H^4(3 a^2 + 2 r_H^2)~~~~~;
P_4= a^6 (24 a^2 + 95 r_H^2 )+ a^2 r_H^4(142 a^2 + 71 r_H^2)~,
\end{eqnarray}
and $\textrm{Elliptic}K[x]$ and $\textrm{Elliptic}E[x]$ are complete Elliptic integrals of first and second kind respectively. The expressions are given by,
\begin{eqnarray}
\textrm{Elliptic}E[x]= \int_0^{\pi/2} (1-x \sin^2\theta)^{(-1/2)} d\theta;~~~ \textrm{Elliptic}K[x]= \int_0^{\pi/2} (1-x \sin^2\theta)^{(1/2)} d\theta.
\end{eqnarray}

\end{widetext}

\end{document}